\documentclass[12pt]{article}
\usepackage{graphicx}
\usepackage{xcolor}
\usepackage[english]{babel}
\usepackage[latin1]{inputenc}
\usepackage{amsmath,amssymb}
\usepackage{bm}
\date{}
\begin{document}

\title{\bf CONNECTIVITY AND RIGIDITY \\ IN BOROSILICATE GLASSES }

\maketitle

\author{Richard Kerner}$^*$,  \\

$*$  {Address : 
{\small LPTMC, Sorbonne-Universit\'e - CNRS UMR 7600 , \\ Tour 13-23, 5-\`eme , Boite 121,
4 Place Jussieu, 75005 Paris, France \\
Tel.: +33 1 44 27 72 98, \,  Fax: +33 1 44 27 51 00,} \\  {\small email : richard.kerner@sorbonne-universite.fr}}

{\abstract{We present a structural analysis of glasses formed by mix of $SiO_2$ and $B_2O_3$ glass formers
with soda and lime modifiers ($Na_2 O$ and $CaO$), which provide a good testing ground for Stochastic Agglomeration Theory.
With local structural units properly identified, we can reproduce the one-parameter glass transition temperature $T_g (z)$ curve
for the family of $(0.75-z) \; SiO_2 + 0.15 \; Na_2O + 0.10 \; CaO + z \; B_2O_3$ glasses 
studied experimentally by Smedskjaer {\it et al.} }}

\section{Introduction} 

Our aim is to check the validity of simple mathematical models of glass transition by applying them to the soda-lime borosilicates studied by 
Smedskjaer, Mauro {\it et al} \cite{Smedskjaer}. The model we shall use here is known as the ``Stochastic Agglomeration Theory'' (see \cite{Kerner1995}, \cite{Kernerbook}).
It predicts correctly the dependence of the glass transition temperature $T_g$ on the modifier's concentration in binary or ternary covalent glasses,
like $Se_{(1-x)} As_x$, $Se_{(1-x)} Ge_x$, etc.

The data on the soda-lime borosilicates obtained by Smedskjaer, Mauro et al. provide an excellent ground for testing the predictive power 
of stochastic agglomeration models of glass transition, in particular the dependence of the glass transition temperature $T_g$ on the average connectivity and rigidity.
A four-component soda-lime borosilicate's composition depends on three variables:
$(1-x-y-z) SiO_2 + x \; Na_2 O + y\; CaO + z \; B_2O_3,$ 
and requires a lot of effort in order to get an adequate picture of how chemical composition influences its thermomechanical properties.
This is why a more modest task was chosen: to reduce the problem to only one variable $z$ (the $B_2O_3$ molar content) while keeping the proportions
of the remaining three components always the same, i.e. those of the most common window glass recipe. 

\begin{equation}
(0.75 - z) \; SiO_2 + 0.15 \; Na_2 O + 0.10 \; CaO + z \; B_2O_3.
\label{windowborox}
\end{equation}

This choice being made, the problem can be treated as in the case of {\it binary glasses}, the window glass considered as the glass-former, and the $B_2O_3$ component 
playing the role of a ``modifier'', although it is a good glass former by itself. In order to avoid misunderstanding in what follows, we shall refer to it as ``additive''.  

It turns out that the main formulas of the stochastic agglomeration model (see e.g. \cite{DMDSRKMM}, \cite{BarrioNaumis}), in particular the dependence 
of the glass transition temperature on modifier's concentration $T_g(z)$ and its initial slope at $z=0$ are in perfect agreement with experimental data. 

\section{Glass forming and connectivity}

The main idea of the constraint theory proposed by J.C. Phillips is that the best glass forming conditions are realized by random atomic networks which are {\it isostatic}.
This concept was conceived by Maxwell who introduced a simple method of evaluation of both angular (bending) and linear (stretching) constraints in solid mechanical structures,
which can be generalized for atomic networks as well.  
A network is called ``isostatic'' when the average number of constraints per atom, $\frac{N_c}{N_a}$, is equal to $3$, the number 
of degrees of freedom of a point in three dimensions. When $\frac{N_c}{N_a} < 3$, the network is underconstrained, or ``floppy'', and when 
$\frac{N_c}{N_a} > 3$, the network is overconstrained.  

As proposed by J.C. Phillips, let us add up the angular constraints imposing the direction of bonds 
stemming from a given atom, and the linear constraints imposing bonds' length. The bonds being shared by two atoms, each linear constraint is counted with
factor $\frac{1}{2}$, i.e. $\frac{m}{2}$ for an $m$-valenced atom. 

A two-valenced atom $(m=2)$ with two bonds keeping a constant angle imposed by interatomic potentials contributes
with one angular constraint, needed to fix the angle between the two bonds pointing out towards its neighbors. 
Adding a new bond requires two additional angles to be fixed  with respect to the bonds already in place, thus adding two extra constraints.
Therefore the number of angular constraints fixing bond directions stemming from an $m$-valenced atom is given by the formula 
\begin{equation}
N_{\alpha} = 2m - 3.
\label{Nalpha}
\end{equation} 
In some cases $m=1$ also makes sense, e.g. when an alkali ion $Na^+$ ``eats out'' one of the bonds in the $SiO_2$ network.
A three valenced atom adds $\frac{3}{2}$ linear constraints to the total count, and $3$ angular constraints fixing the angles between its three bonds. 

The formula giving the total number of constraints added by an $m$-coordinate atom takes into account the 
$\frac{m}{2}$  linear constraints, coming from the $m$ bonds shared in half with its neighbors in the network, 
and $2m - 3$  angular constraints fixing the angles between $m$ unit vectors. The general formula is therefore 

\begin{equation}
N_c (m) = \frac{m}{2} + 2 m - 3.
\label{Nc}
\end{equation}

The isostaticity condition is readily translated into the optimal average coordination number: it is enough to require, according to (\ref{Nc}),
\begin{equation}
N_c (m) = \frac{m}{2} + 2 m - 3 = 3, \; \; \; \rightarrow \; \; \; m = \frac{12}{5} = 2.4.
\label{magic}
\end{equation}
Surprisingly, the same number appears in graphs known as {\it Bethe lattices}. The simplest case of which is a homogeneous dendritic structure made of identical 
$m$-valenced star-like objects, linked by standard length bonds to its neighbors. Starting from the first $m$-coordinate ``seed'', one adds next $m$ atoms
forming the first layer neighbors, each of whom provides $(m-1)$ free bonds available for the next atoms, and so on. If the formation of closed rings is excluded,
the number of new open bonds created in consecutive layers will grow as $(m-1)^k$, whereas in three dimensions the surface of consecutive spheres on which the new bonds 
point out grows at most as $(k d)^2$; where $d$ is the average bond's length.   

\begin{figure}[hbt]
\centering 
\includegraphics[width=2.8cm, height=2.8cm]{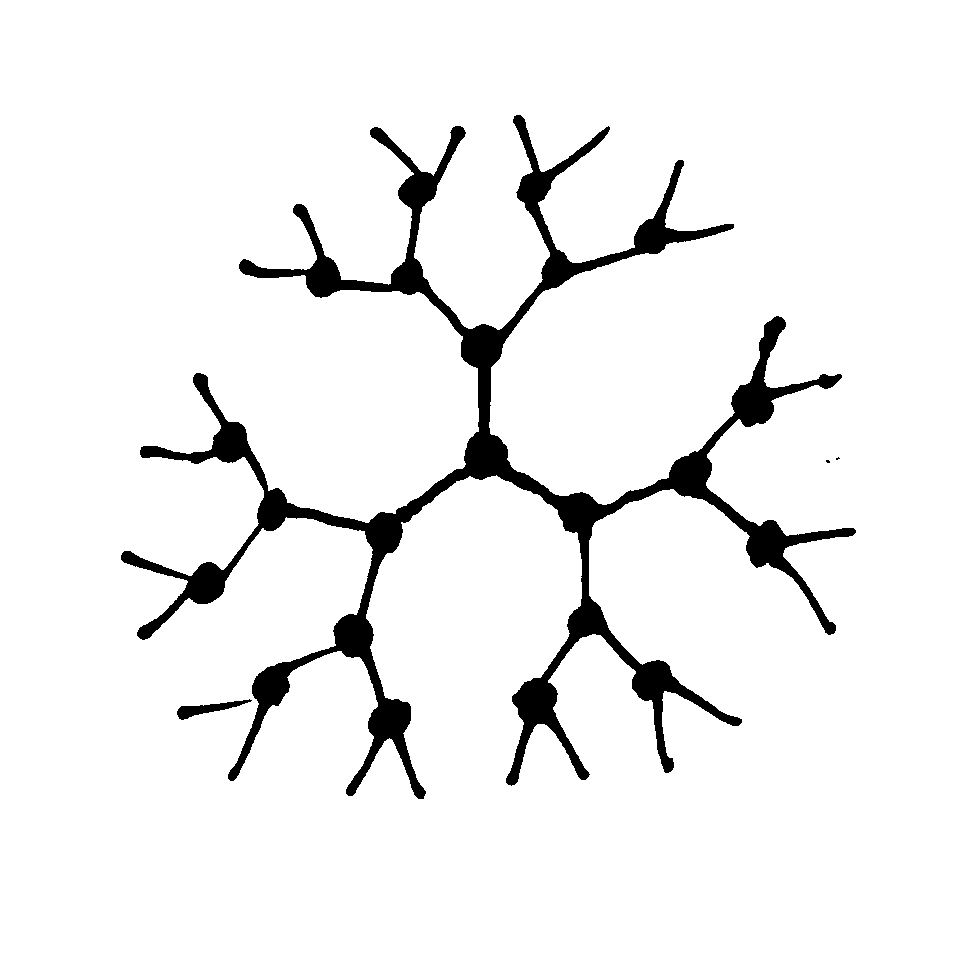}
\hskip 0.4cm
\includegraphics[width=2.8cm, height=2.8cm]{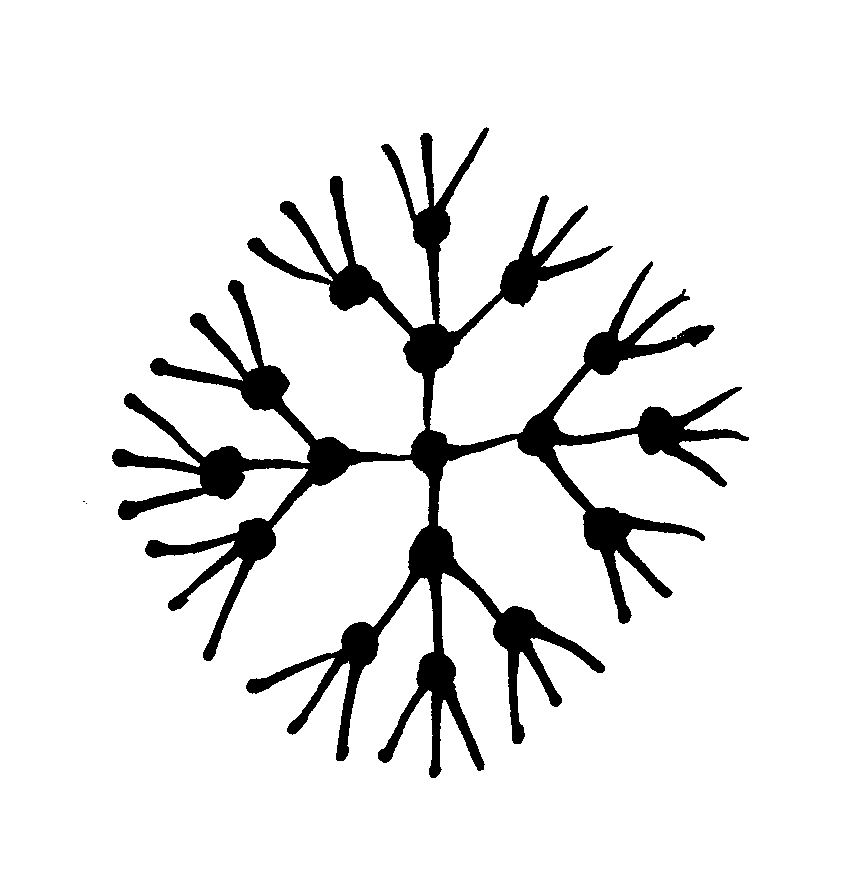}
\hskip 0.4cm
\includegraphics[width=2.8cm, height=2.8cm]{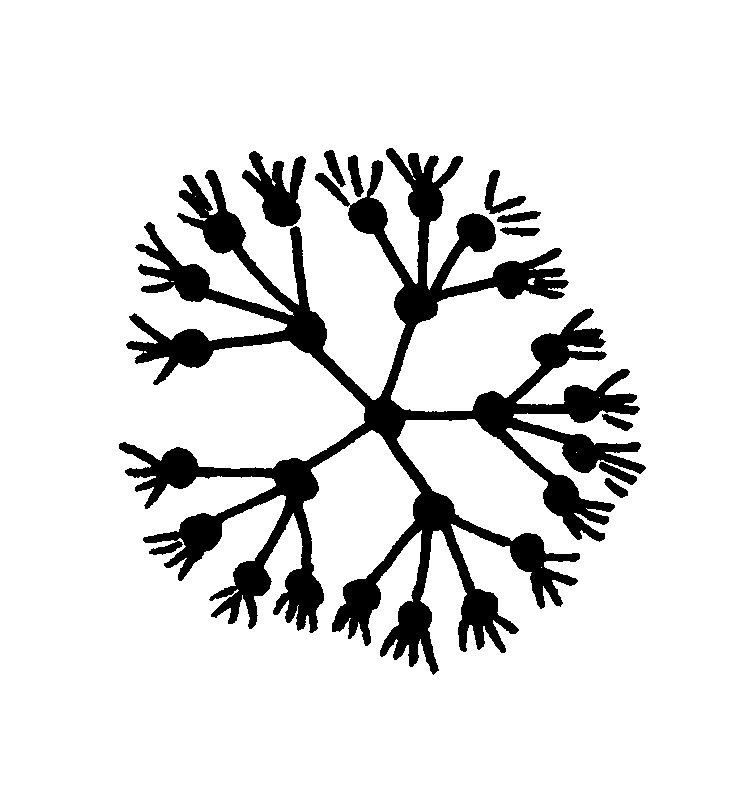}
\label{fig:Bethelattices}
\caption{Three Bethe lattices made of $3$, $4$ and $5$-coordinated atoms}
\end{figure}
A purely dendritic structure can be analyzed by looking at the successive layers
surrounding  a randomly chosen  atom.  In order to stabilize density on the successive layers of growing cluster, 
the stoichiometric ratio $2:3$ (e.g. $As_2 Se_3$) is enough. The average coordination number is then equal to 
\begin{equation}               
<N_c>= (2 \times 3 + 3 \times 2)/(2+3) = 12/5 = 2.4,
\label{As2Se3}
\end{equation}
\begin{figure}[hbt]
\centering 
\includegraphics[width=2.8cm, height=2.8cm]{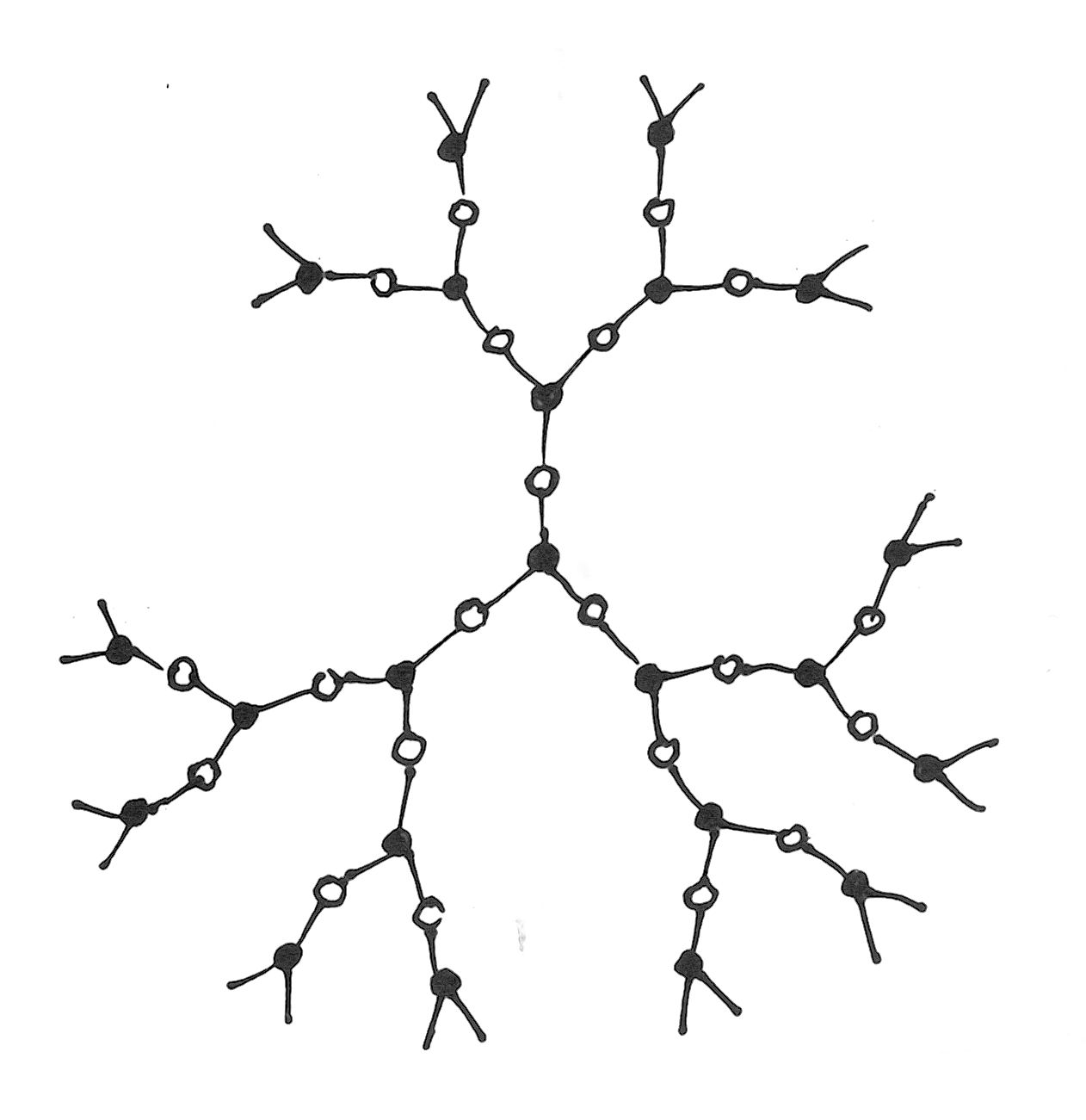}
\hskip 0.4cm
\includegraphics[width=3.3cm, height=3.3cm]{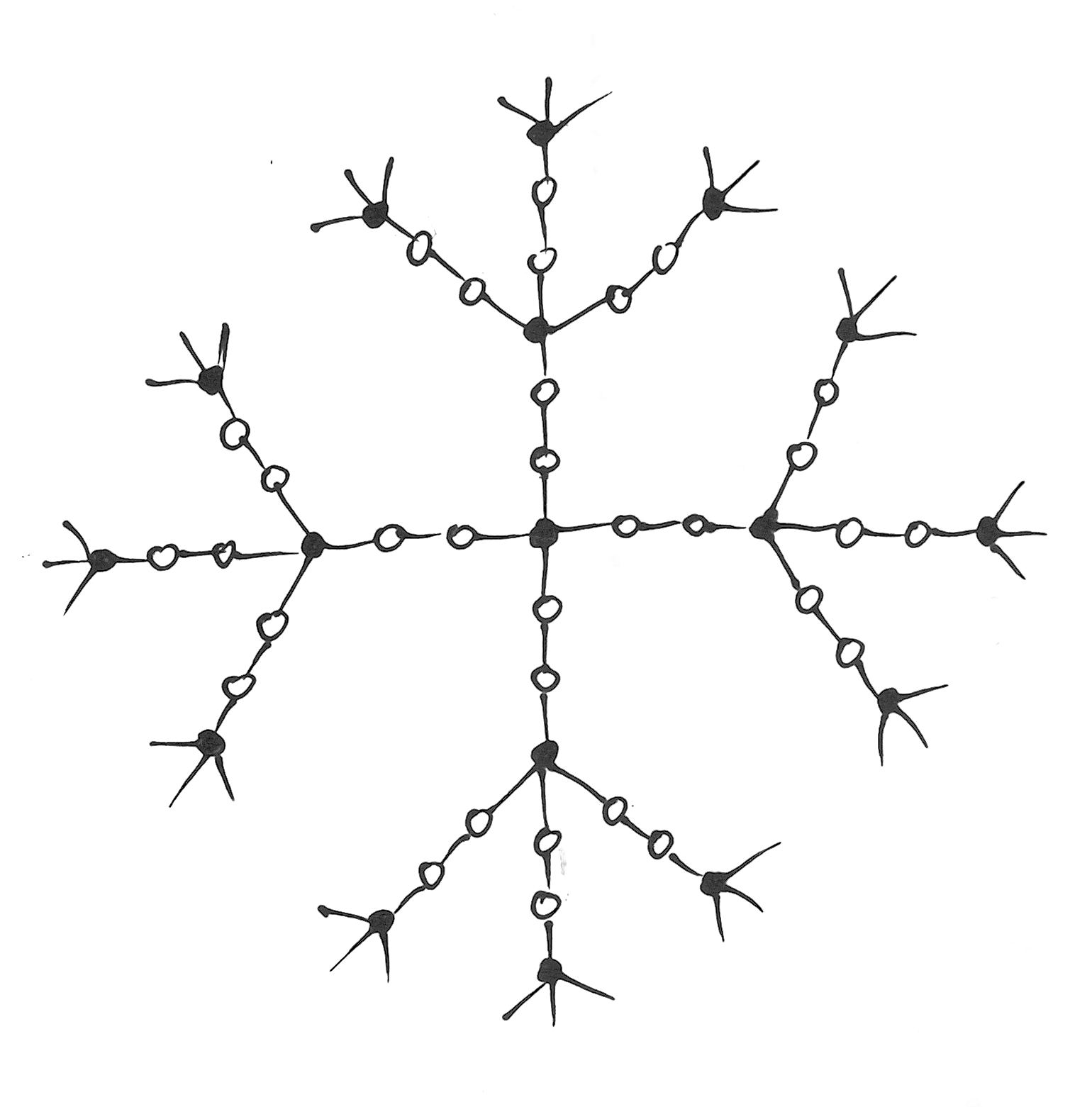}
\hskip 0.4cm
\includegraphics[width=3.3cm, height=3.3cm]{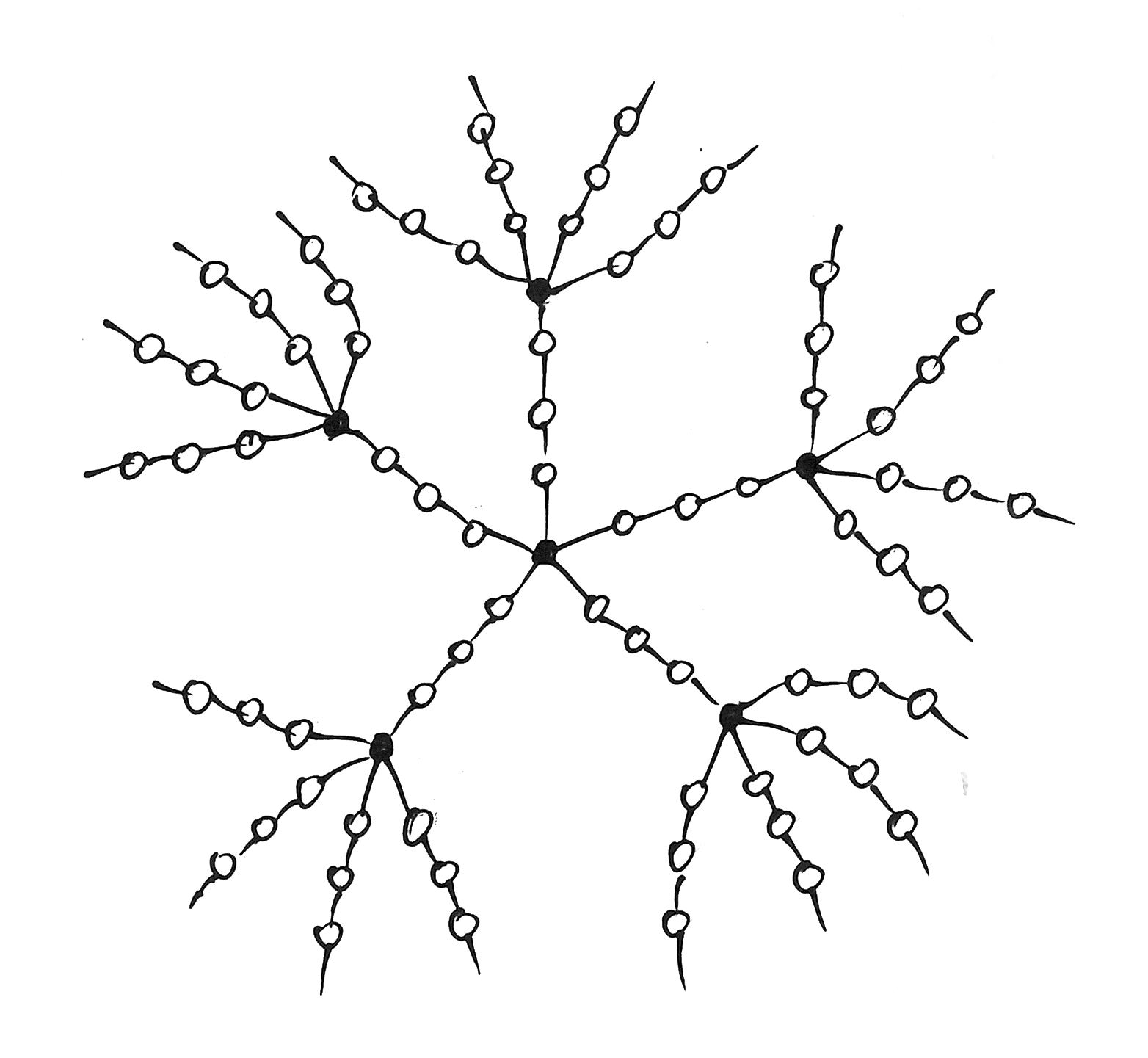}
\label{fig:BethelatticesB}
\caption{The same Bethe lattices diluted with $2$-valenced atoms.}
\end{figure}
In the case of four-coordinate atoms ($Si, Ge..$) interconnected by atoms of valence two ($S, Se$) the distances between the four-valenced atoms 
have to increase even more. This is done by intercalating an extra $S$ or $Se$ atom in between, i.e. rescaling the bond length by factor $2$, 
thus creating a network with the stoiichiometry $Ge Se_4$.  Now the average coordination will be:
\begin{equation}                     
<N_c> = (1 \times 4 + 2 \times 4)/(1+4) = 12/5 = 2.4 
\label{BethelatticesC}
\end{equation}

The five-coordinate atoms (like phosphorus $P$) need as much as three two-valenced $Se$ atoms to dilute the density of the corresponding Bethe lattice.
The count is then 
\begin{equation}
<N_c> = (1 \times 5 + 7.5 \times 2)/(1 + 7.5) = 2.353
\label{BethelatticeD}
\end{equation}
which is still very close to the magic average coordination number $2.4$. 

\section{Window glass - the optimal composition}

The study of soda-lime borosilicate glasses presents a mathematical challenge of higher order than in the case of alkali-silicates or alkali-borates,
described by a single parameter (the alkali modifier's content), or the common window glass, which is a lime-soda silicate parametrized by two variables,
the alkali and lime content. The soda-lime borosilicates contain four components, and their characterization needs three parameters, as shown in
the corresponding chemical formula: $(Si0_2)_{(1-x-y-z)} (Na_2O)_x (CaO)_y (B_2 O_3)_z $.

The composition of glasses studied by Smedskjaer et al. in \cite{Smedskjaer} was restricted in a way that left only one free parameter $z$, while
the concentrations of $Na_2O$ and $CaO$ were fixed, their values characteristic for the most common window glass, with $x = 0.15$ and $y=0.10$:
\begin{equation}
0.75 \; SiO_2 + 0.15 \; Na_2O + 0.10 \; CaO
\label{windowglass}
\end{equation}
This particular composition, known since the Antiquity, can be derived theoretically, using simple principles of topological nature (see \cite{RKJCP2000}, \cite{JCPRK2008}).
Let us give here a concise derivation. Consider a soda-lime silicate glass with molar composition given by 
$$(1-x-y) SiO_2 + x Na_2O + y CaO$$
where $x$ is the molar concentration of $Na_2O$, $y$ is the molar concentration of $CaO$ (lime). 
To ensure that the network is {\it isostatic}, it should display the optimal average coordination number  $<m> =2.4$. Therefore the unknown molar concentrations 
should satisfy the following linear equation: 
\begin{equation}
<m_{SiO_2}> (1-x-y) + <m_{Na_2O}> x + <m_{CaO}> y = \frac{12}{5} = 2.4
\label{Zeroxy}
\end{equation}
We easily find out the average coordination numbers of particular compounds: 
$$<m_{SiO_2}> =  \frac{8}{3}, \; \; \; <m_{Na_2O}> = \frac{4}{3}, \; \; \; <m_{CaO}>= \frac{4}{2}.$$
The explicit equation becomes : 
\begin{equation}
\frac{8}{3} (1-x-y) + \frac{4}{3} x + \frac{4}{2} y = \frac{12}{5},
\label{Firstxy}
\end{equation} 
resulting in a simple relation: 
\begin{equation}
2 x + y = 0.4
\label{xysimple}
\end{equation}
This fixes a straight line in our two-dimensional parameter space $(x,y)$.
A second equation is needed to determine the optimal values of $x$ and $y$. The average coordination number
takes into account only the immediate neighborhood of each atom, which is often insufficient to distinguish between crystalline or glassy
structures, especially in covalent chalcogenide glasses or a pure $SiO_2$. Radial distribution functions display the same first peak
in both cases, crystal or glass, the only difference being in the shape of the first peak, very sharp in crystals, and more like
a Gaussian in glasses. But from the second peak on, corresponding to the second and farther neighbors, the picture gets totally
different; consequent sharp delta-function like peaks in crystals, and diluted and barely visible local maxima in glasses, whose radial
distribution function tend to display a uniform behavior like in a perfect gas. 

When the network forming atoms display essentially different physical and chemical properties, the average coordination number does not provide 
sufficient information. It seems obvious that no less essential information can be learned from the medium-range atomic configurations, and the most important
feature is the distribution of rings, which determines the medium-range order. In a pure $SiO_2$ crystalline structures only six-fold rings are
present, called ``a chair'' and ``a boat'', each silicon atom being a common vertex of six rings. 

In the amorphous $SiO_2$ the six-fold rings still dominate, but the Raman spectroscopy and other experimental methods reveal the presence of
$5$ and $7$ fold rings, and even a certain amount of $4$ and $3$-fold puckered rings (see \cite{Galeener}, \cite{Geissberger}, \cite{Kernerrings}). However, the number
of minimal rings built around each $Si$ atom remains the same, i.e. equal to $6$, as follows from simple combinatorics: it is the number of diffeent
pairs of bonds stemming from a $4$-coordinate center, and this is equal to $C^4_2 = \frac{4!}{2! (4-2)!} = 6$.

In soda-lime silicate glasses three local structure units dominate (\cite{Vedishcheva}):  
\vskip 0.1cm
{\centerline{ $  Q^4 = Si {\O{}}_4, \; \; \; Q^3 = Si {\O{}}_3 O Na^+, \; \;  2Q^3 = Si {\O{}}_3 O Ca O Si {\O{}}_3 $ }}

\begin{figure}[hbt]
\centering 
\includegraphics[width=1.5cm, height=1.5cm]{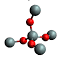}
\hskip 1.4cm
\includegraphics[width=1.5cm, height=1.5cm]{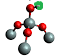}
\hskip 1.3cm
\includegraphics[width=2.5cm, height=1.5cm]{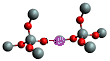}
\label{fig:Q3Q4Q6}
\caption{\small $  Q^4 = Si {\O{}}_4, \; \; \; \; \; \; Q^3 = Si {\O{}}_3 O Na^+, \; \; \;   2Q^3 = Si {\O{}}_3 O Ca O Si {\O{}}_3 $ } 
\end{figure}

\begin{figure}[hbt]
\centering 
\includegraphics[width=11cm, height=3.4cm]{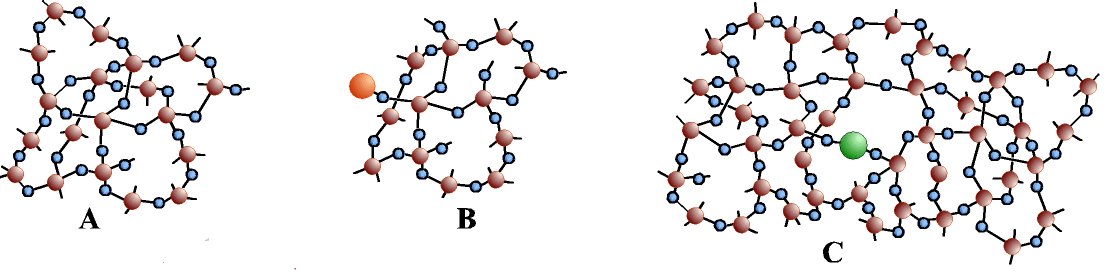}
\label{fig:Rings}
\caption{\small The pairs of bonds pointing towards closest neighbors give rise to closed rings. A simple combinatorics gives their numbers: 
$6$ distinct rings around each $Q^4$ unit (A), three rings only around each $Q^3$ unit (B), and as much as $15$ rings around each pair of silicon atoms 
``zipped'' together by a $Ca$ atom (C).} 
 
\end{figure}

In crystalline forms of pure $SiO_2$ there are always six $6$-folded rings around each $Q^4$ structural unit, with which the network is formed. 
In order to guarantee maximal homogeneity of the network modified by the presence of alkali and lime molecules, let us postulate that
the AVERAGE NUMBER of rings around the new structural units remains as close to $6$ as possible.

Now, each molecule of $Na_2O$ neutralizes two oxygen bonds by sticking an $Na^+$ ion
to one of the oxygens, and breaking its connection to a next $Si$ atom, thus transforming TWO $Q^4$ units into 
two $Q^3$ units. The number of rings becomes $2 \times 3$ instead of $2 \times 6$, suppressing $6$ rings out of the total number. 
A $CaO$ molecule produces ONE ``zipped'' configuration, with $6$
outward pointing bonds, which give rise to rings - an excess of $9$ rings as compared with the average $6$. 
In order to keep the balance, we must have  
\begin{equation}
 x \times 6 = y \times 9 \; \; \; \rightarrow 2 x = 3 y,
\label{ringequation}
\end{equation}
Now we can solve the system, evoking the first equation $2 x + y = 0.4$. 

The result, which is the conjunction of two conditions, to have an isostatic network with the average coordination number $2.4$,
and to ensure maximal homogeneity on the medium- range level (average number of rings around each structural unit) is the best
recipe for {\it window glass}, known since the Middle Ages: 

\begin{equation}
75 \% \; SiO_2, \; \; \; 15 \% \; Na_2O , \; \; \; 10 \% \; CaO.
\label{composition}
\end{equation}

\section{The agglomeration model and $T_g$}

The glass forming process, and the glass transition in particular, can be successfully modelled as an agglomeration process in overcolled liquid melts.
 Let us present it in a nutshell. 

A slightly overcooled liquid contains clusters of various sizes which agglomerate progressively when the temperature goes slowly down. In the case of
covalent glass formers, such as $As_2 Se_3$, $Ge Se_2$, or $As-Ge-Se$ mixtures, the agglomeration starts between atoms, and the relative frequency of
creating clusters depends crucially on the valence (coordination number) and energy landscape providing potential energy costs for creating a particular bond.

\begin{figure}[hbt]
\centering 
\includegraphics[width=6cm, height=3.4cm]{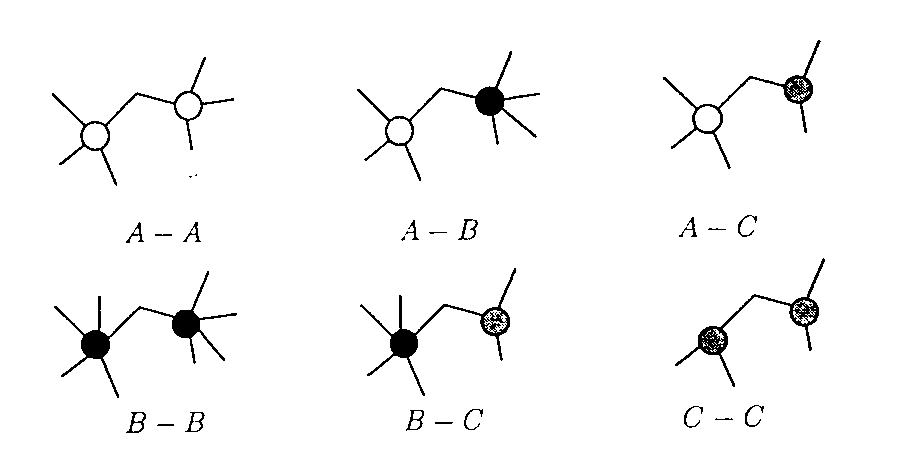}
\label{fig:ABCDoublets}
\caption{Six different doublets formed with three types of atoms}
\end{figure}

The model is even simpler when only two types of agglomerating items are present. 

Let us therefore suppose the presence of two types of building blocks (single atoms, or small entities made from several atoms) with $m_{\rm A}$
and $m_{\rm B}$ denoting the number of their free valencies. This means that each block A will have  $m_{\rm A}$ neighbors in the resulting
structure and each B-block will be surrounded by $m_{\rm B}$ immediate neighbors.  

Let the concentration of B-blocks be $c$, and the concentration of the A-blocks be $1-c$. These probabilities
represent just the number of respective blocks divided by number of all blocks present in the network. 

We describe the the probabilities with the help of two factors:
\begin{itemize}
\item the statistical weight for each agglomeration process; that is, the number of ways leading to the same result;
\item the Boltzmann factor taking into account the energy barrier of forming a certain type of bond.
\end{itemize}
\vskip 0.2cm
Let us assume that the energy cost of forming a bond between two A-blocks is $E_{\mathrm{AA}}$\,, between an A-block and a B-block $E_{\mathrm{AB}}$ \,, 
and between two blocks of the type B it is $E_{\mathrm{BB}}$. 
The ultimate global network forming process occurs at a given $T_{\mathrm{g}}$, which depends on the concentration $c$.

In order to find the concentrations of different types of building blocks in the resulting glass structure, we should express the
probabilities of all different types of bonds (or doublets, i.e. pairs of adjoining blocks).  
We take into account the number of ways they can stick to each other (depending on the number of free bonds of two blocks about to
agglomerate), the concentration of a given type of block, and the energy barrier represented by the Boltzmann factor ${\rm e}^{-E/(k\,T)}$,
where $E$ is the energy, $k$ the Boltzmann constant and $T$ the absolute temperature. 

The probabilities of finding doublets A-A, A-B, and B-B, forming at $T_{\mathrm{g}}$ become:
\begin{equation}
\label{AA} 
p_{\mathrm{AA}} =\frac1{Q}\,m^2_{\mathrm{A}}\,(1-c)^2\,
{\rm e}^{-E_{\mathrm{AA}}/(k\,T_{\mathrm{g}})}, 
\end{equation}
\begin{equation}
\label{AB} p_{\mathrm{AB}}
=\frac1{Q}\,2\,m_{\mathrm{A}}\,m_{\mathrm{B}}
\,(1-c)\,c\,{\rm e}^{-E_{\mathrm{AB}}/(k\,T_{\mathrm{g}})},
\end{equation}
\begin{equation}
\label{BB} p_{\,\mathrm{BB}} =\frac1{Q}\,m^2_{\mathrm{B}}\,c^2\,
{\rm e}^{-E_{\mathrm{BB}}/(k\,T_{\mathrm{g}})},
\end{equation}
where the normalizing factor $Q$ is the sum 
\begin{equation}
m^2_{\mathrm{A}}\,(1\!-\!c)^2\, {\rm
e}^{-E_{\mathrm{AA}}/(k\,T_{\mathrm{g}})}+
2\,m_{\mathrm{A}}\,m_{\mathrm{B}} \,(1\!-\!c)\,c\,{\rm
e}^{-E_{\mathrm{AB}}/(k\,T_{\mathrm{g}})}+m^2_{\mathrm{B}}\,c^2\,
{\rm e}^{-E_{\mathrm{BB}}/(k\,T_{\mathrm{g}})}. 
\end{equation}
The probability of finding a block $B$ in the collection of all doublets can be expressed as 
\begin{equation}
p_{\,\rm B}=\frac{1}{2 } \, \left[ p_{\rm AB}+ 2 p_{\rm \,BB}\right] \,.
\label{probB}
\end{equation}
(We count the probability $p_{AB}$ only once, because there is only one item $B$ in the doublet $AB$, and we count twice the probability $p_{BB}$ because
there are two items $B$ per $BB$ doublet; finally, the sum is divided by $2$ in order to normalize the new probabilities to $1$.).

Now we require that the probability distribution among the doublets is the same as among the single building blocks, 
which can be interpreted as minimizing statistical fluctuations. This leads to the following equation: 
\begin{equation}
\frac{1}{2} \left( p_{AB} + 2 p_{BB} \right) = c
\label{equilibc}
\end{equation}
which yields, after multiplying both sides by $Q$
\begin{equation}
c (1-c) \left( m_{\mathrm{A}}m_{\mathrm{B}}(1-2\,c)-m^2_{\mathrm{A}}(1-c)\,\xi
+m^2_{\mathrm{B}}c\,\mu \right) =0 \,,
\label{equationc}
\end{equation}

The third-order algebraic equation for $c$ has always two solutions $c = 0$ and $c = 1$; depending on the values of the Boltzmann and statistical 
weight factors {\it may have} or {\it have not} a third solution contained between $0$ and $1$, that we shall denote by 
$c_{am}$. The two situations are displayed in the figure ($6$): 
\begin{figure}
\centering
\includegraphics[width=6cm, height=1cm]{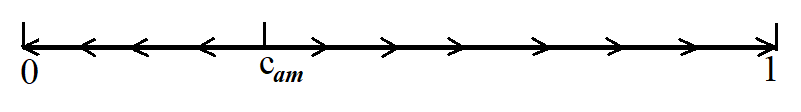}
\hskip 0.3cm
\includegraphics[width=6cm, height=1cm]{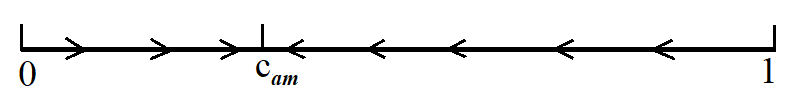}
\caption{Two diagrams representing the behavior of the system: 
a) with stable solutions at $c=0$ and $c=1$, b) with stable solution at $c = c_{am}$ }
\label{fig:Twosolutions}
\end{figure}
Two obvious solutions $c=0$ and $c=1$ correspond to pure $A$ or $B$
compositions; the glass is formed as a mix of the two, given by the third solution.  
With the condition $p_{\,\rm B}=c$ and \ref{AA}, \ref{AB} and \ref{BB}, we get the third solution,

\begin{equation}
\label{fii}
\Phi
(c,T_{\mathrm{g}})=m_{\mathrm{A}}m_{\mathrm{B}}(1-2\,c)-m^2_{\mathrm{A}}(1-c)\,\xi
+m^2_{\mathrm{B}}c\,\mu=0\,,
\end{equation}
which gives implicitly the dependence of the glass transition temperature on the concentration $c$. The dependence on the
temperature is there hidden in the parameters
\begin{equation}
\xi(T_{\mathrm{g}})={\rm e}^{(E_{\rm AB}-E_{\rm AA})/(k\,T_{\mathrm{g}})} \; \; {\rm  and} \; \;  \mu(T_{\mathrm{g}})=
{\rm e}^{(E_{\rm AB}-E_{\rm BB})/(k\,T_{\mathrm{g}})}.
\label{ximu1}
\end{equation}

Introducing explicitly the Boltzmann factors, $ e^{- \alpha_i} = e^{- \frac{E_i}{kT}}$,  we get the following simplified expression for $c_{am}$:
\begin{equation}
c_{am} = \frac{ m_A m_B e^{- \alpha_2} - m_B^2 e^{- \alpha_3}}{ 2 m_A m_B 
e^{- \alpha_2} - m_B^2 e^{- \alpha_3} - m_A^2 e^{- \alpha_1}} .
\end{equation}
\noindent
The condition for $c_{am}$ to be contained in the interval $[0, 1]$ is
equivalent to:
\begin{equation}
m_A e^{- \alpha_2} > m_B e^{- \alpha_3} \ \ {\rm and} \ \ m_B e^{- \alpha_2} >
m_A e^{- \alpha_1} .
\end{equation}

(Note that this condition will remain valid if we exchange simultaneously $m_A$ with $m_B$ and $\alpha_1$ with $\alpha_3$; but this amounts to re-define
the roles of the ``basis" and ``modified" states, which is one of the symmetries of our approach in the case of binary systems.)

A good glass former should display the amorphisation tendency independently of the modifier rate; this means that the solution $c_{am}$ has a continuous
limit when $x \rightarrow 1$. In this case, the condition reduces to
\begin{equation}
m_B e^{- \alpha_2} = m_A e^{- \alpha_1} .
\end{equation}
Introducing explicitly the Boltzmann factors, $ e^{- \alpha_i} =
e^{- \frac{E_i}{kT}}$, we obtain:
\begin{equation}
\frac{E_2 - E_1}{kT} = \ln \, (\frac{m_B}{m_A}) \; \; {\rightarrow} \; \; (E_2 - E_1) - kT \; (\ln m_B - \ln m_A) = 0. 
\label{fundamental}
\end{equation}
The formula (\ref{fundamental}) can be interpreted as condition of minimal variation of {\it free energy} at constant temperature, which in classical thermodynamics
reads as $\delta F = \delta U - T \delta S$. The energy variation is defined by the difference $E_2 - E_1$, while the entropy variation is reduced to $\ln m_B - \ln m_A$.
The originality here is that what we compare here are not {\it states} but {\it agglomeration processes}.  

Let us analyze this solution for low values of concentration $c$. In the limit $c \rightarrow 0$, we get, 
with the help of the corresponding transition temperature $T_{\mathrm{g}0}=T_{\mathrm{g}}|_{c=0}$\,, 
\begin{equation}
\label{xic0}
m_{\rm A}\,\xi \Big|_{T_{\mathrm{g}}=T_{\mathrm{g}0}}\!- m_{\rm
B}=0 \qquad {\rm or}\qquad {\rm e}^{(E_{\rm AB}-E_{\rm
AA})/(k\,T_{\mathrm{g}0})}= \frac{m_{\rm B}}{m_{\rm A}} \, .
\end{equation}

The initial slope of the transition temperature curve can be then obtained from \ref{fii} and \ref{xic0} as follows 
\begin{equation}
\label{slope}
\left.\frac{{\rm d}T_{\mathrm{g}}}{{\rm d}c}\right|_{c=0}=
\frac{T_{\mathrm{g}0}} {\ln\left(\frac{m_{\rm B}}{m_{\rm A}}\right)}
\left(1-\frac{m_{\rm B}}{m_{\rm A}}\,\mu
\Big|_{T_{\mathrm{g}}=T_{\mathrm{g}0}}\right) .
\end{equation}

With $\mu=0$ we get an even simpler formula 
\begin{equation}
\label{elegant}
\left.\frac{\mathrm{d}T_{\mathrm{g}}}{\mathrm{d}c}\right|_{c=0}=\frac{T_{\mathrm{g}0}}{\ln
\left(\frac{m_{\mathrm{B}}}{m_{\mathrm{A}}}\right)}.
\end{equation}
It shows that the initial slope of the $T_{\mathrm{g}}(x)$ curve in covalent binary glasses 
depends only on the ratio of the valencies of the glass-former and the modifier. It works perfectly for covalent network glasses, 
where is equal to $2$ for $\mathrm{Ge}_x\mathrm{Se}_{(1-x)}$ glass, and to $3/2$ for $\mathrm{As}_x\mathrm{Se}_{(1-x)}$ glass. 

A linear shape of the curve $T_g(z)$ is characteristic for binary glasses whose
 components have the same coordination number, i.e. when $m_A = m_B$, and both are good glass formers. The formula 
(\ref{elegant}) becomes singular, because then
$\ln(m_B/m_A) = 0$. But it is an approximation valid for $c \rightarrow 0$, when the energy barrier $E_{BB}$ did not appear.
It turns out that the full formula can be remastered taking into account both ends of the range $0 \leq c \leq 1$. 
Let us recall the abbreviated notation 
\begin{equation}
\xi = e^{\eta -\epsilon} = e^{\frac{E_{AB}-E_{AA}}{kT}} \, \ \ \,
{\rm  and} \, \ \ \,  \mu = e^{\eta - \alpha}= e^{\frac{E_{AB}-E_{BB}}{kT}}.
\label{ximu2}
\end{equation}

\indent
We can rewrite our minimal fluctuation condition in a symmetric manner, invariant
with respect to the simultaneous substitution $m_A \leftrightarrow m_B$,
$c \leftrightarrow (1-c)$ and $\xi \leftrightarrow \mu$:
\begin{equation}
c(1-c) \, [ (1-c) \, (1 - \frac{m_A}{m_B} \, \xi ) -
c \, ( 1 - \frac{m_B}{m_A} \, \mu  ) ] = 0 .
\label{symcondition1}
\end{equation}
Obviously, the ``pure states" $c = 0$ or $c = 1$ represent stationary
solutions of (\ref{symcondition1}) and can be factorized out. The non-trivial
condition for the glass forming is thus
\begin{equation}
(1-c) \, [ 1 - \frac{m_A}{m_B} \, \xi ] - c \, [1 - \frac{m_B}{m_A} \, \mu ]  = 0 .
\label{symcondition2}
\end{equation}
Now, using the limit conditions at $c \rightarrow 0,  \, \ \ T_g = T_{g0}$
and $c \rightarrow 1 , \,  \ \ T_g = T_{g1}$, and
introducing the generalized Boltzmann factors with the energy barriers
for corresponding bond creations as $E_{AA}, \, E_{AB} \,$ and $E_{BB}$,
we can write
\begin{equation}
E_{AB} - E_{AA} = k \, T_{g0} \, ln ( \frac{m_B}{m_A}) \, ,
\, \ \ \,
E_{AB} - E_{BB} = k \, T_{g1} \, ln ( \frac{m_A}{m_B} ) , 
\label{energies3}
\end{equation}
so that the expressions $\xi$ and $\mu$ at the arbitrary temperature $T$ become
$$
\xi (T) = e^{\frac{E_{AB} - E_{AA}}{ T_{g0}}  \cdot \frac{T_{g0}}{T}}
= ( \frac{m_B}{m_A} )^{\frac{T_{g0}}{T}} \, ;$$ 
\begin{equation}
\mu  (T) = e^{\frac{E_{AB} - E_{BB}}{ T_{g1}}  \cdot \frac{T_{g1}}{T}}
= ( \frac{m_A}{m_B} )^{\frac{T_{g1}}{T}} . 
\label{xi}
\end{equation}
Substituting these expressions into  (\ref{slope}) and
taking the limit $c \rightarrow 0$, we get
\begin{equation}
\frac{d T_g}{dc}\mid_{c=0} = \frac{ T_{g0} [1 - (\frac{m_B}{m_A})^{\frac{ T_{g0} -
T_{g1}}{T_{g0}}} ] }{ln(\frac{m_B}{m_A})} .
\label{unislope}
\end{equation}
It is easy to see now that even when $m_A = m_B$, this formula has
a well defined limit. Indeed, if we first set $\frac{m_B}{m_A} =
1 + \epsilon $, and then develop the numerator and the denominator of the
above equation in powers of $\epsilon$, then in the limit when $\epsilon
\rightarrow 0 $ we arrive at a simple linear dependence which is
in agreement with common sense and with experiment as well, namely
\begin{equation}
\frac{d T_g}{dc} \mid_{c=0} = T_{g1} - T_{g0} .
\label{Tlinear}
\end{equation}
\indent
This formula is also confirmed by many experiments, e.g. performed on 
selenium-sulfur mixtures (where $m_A = m_B = 2$). We shall show how it works in the case of the soda-lime borosilicates.

\section{Alkali-Lime Borosilicate glass}

The experimental data presented in the article \cite{Smedskjaer} provide an excellent test ground for the stochastic agglomeration model. 
The Table ($1$) displays eight soda-lime borosilicate glasses with varying $SiO_2$ and $B_2O_3$ content, ranging from a pure window
glass up to a pure alkali-lime borate glass, keeping the proportion of $Na_2O$ and $CaO$ modifiers constant. The fraction of four-coordinated boron atoms ($N_4$) 
and the fraction of silicon atoms with four bridging oxygens ($Q^4$) were estimated using MAS and NMR spectroscopy respectively. They suggest the
dispatching of $Na^{+}$ ions between $Si$ and $B$ atoms as their proportion varies.
\hskip 0.2cm

\begin{center}
    \begin{tabular}{ | l | l | l | l | l | l | l |}
    \hline
    $SiO_2$ & $B_2O_3$ & $Na_2O$ & $CaO$ & $N_4$ & $Q^4$ & $T_g (K)$  \\ \hline \hline
    $ 0 $ & $74.1$ & $15.4$ & $10.5$ & $39.9$ & $-$ & 775 \\ \hline
    $12.7$ & $62.0$ & $14.9$ & $10.4$ & $45.3$ & $27$ & 790 \\ \hline						
    $24.9$ & $49.3$ & $15.0$ & $10.8$ & $48.9$ & $33$ & 803 \\ \hline						
    $36.9$ & $38.4$ & $14.1$ & $10.6$ & $53.5$ & $28$ & 813 \\ \hline						
    $51.6$ & $21.9$ & $15.5$ & $11.0$ & $65.5$ & $37$ & 833 \\ \hline						
    $63.8$ & $10.8$ & $14.8$ & $10.6$ & $75.8$ & $31$ & 842 \\ \hline						
    $69.3$ & $4.9 $ & $16.1$ & $9.8$ & $80.8$ & $26$ & 837 \\ \hline						
    $74.8$ & $ 0 $ & $15.2$ & $10.1$ & $ - $ & $28$ & 814 \\ \hline						
\end{tabular}
\label{SMData}
\end{center}
\hskip 0.2cm
\centerline{\small Table $1$ The experimental data presented in $2011$ by Smedskjaer et al. in \cite{Smedskjaer}. }
\vskip 0.3cm
\indent
In order to apply the formulas (\ref{fii}), (\ref{xic0}) and (\ref{slope}) we need to relate these results with the overall distribution of $3$ and $4$ coordinated
clusters which would play the role of similarly coordinated atoms in the original version of the stochastic agglomeration model. In fact, we are tacitly using the ergodicity
hypothesis, replacing comparison between two different glasses with neighboring compositions, with imaginary process transforming melt composition by adding some extra
amount of the $B_2O_3$ additive. 

Let us consider the variation of probability distribution of characteristic local structures present in two different melts
slightly above the glass transition temperature, when we pour some amount of $B_2O_3$ to the window glass, removing the same molar amount of $SiO_2$. 
Initially, in the absence of $B_2 O_3$, we have three types of structural units, as shown in the \ref{fig:Q3Q4Q6}:
$$Q^4 = Si {\O{}}_4, \; \; \; \; Q^3 = Si {\O{}}_3 O Na^+, \; \; \; \; 2Q^3 = Si {\O{}}_3 O Ca O Si {\O{}}_3.$$

In the standard window glass the initial chemical content, in the absence of the $B_2 O_3$ additive, was 
$0.75 \; SiO_2 + 0.15 \; Na_2O + 0.10 \; CaO.$ 
The alkali modifier $Na_2O$ dissociates in the melt, and the $Na^+$ ions break oxygen bonds between neighboring four-coordinate $Q^4$ units,
transforming them into two three-coordinate $Q^3$ units: 
$$ 2 \; Si {\O{}}_4 + Na_2 O = 2 Si {\O{}}_3 O Na^+ , \; \; \; \; {\rm symbolically} \; \; \; \; 2 \; Q^4 + Na_2O = 2 \; Q^3.$$
The most plausible transformation enabling the insertion of $B_2O_3$ into the network dominated by $Q^4$ and $Q^3$  structural units.
In order to keep the connectivity, i.e. to avoid dangling bonds, an $Na^+$ ion migrates from a $Q^3$ unit towards a boron atom,
transforming $Q^3$ into $Q^4$, and $B^3$ into $B^4$. 

\begin{figure}[hbt]
\centering
\includegraphics[width=5.7cm, height=1.7cm]{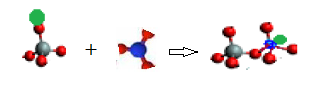}
\caption{A $3$-coordinate $Si-Na^{+}$ unit bonding with a $3$-coordinate Boron yields a new $6$-coordinate structural unit, so that the total connectivity 
($3 + 3 \rightarrow 6$) is conserved.}
\label{fig:SiBreactionA}
\end{figure}

However, at the glass transition temperature bigger clusters dominate the landscape. The analysis of local structures in $B_2 O_3$ in overcooled
liquid melts and in formed glass as well was the subject of many excellent experimental and theoretical investigations, particularly in the papers
by A.C. Hannon et al. (see \cite{Hannon1993}, \cite{Hannon1994}), A.C. Wright and N.M. Vedishcheva (see \cite{Wright1995}, \cite{Wright2016}). 
They reveal the presence of mostly four-coordinate clusters, with or without $Na^+$ ions attached,  shown in the Fig. 8.

\begin{figure}[hbt]
\centering
\includegraphics[width=8cm, height=2.7cm]{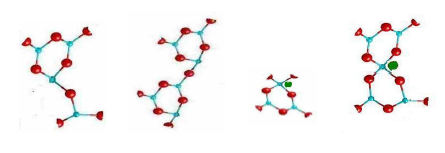}
\label{fig:ABclusters}
\caption{Local $4$-coordinate clusters in $(1-x) B_2O_3 + x Na_2O$}
\end{figure}

The proportion being $\frac{2}{3}$ clusters of the first type (A), and $\frac{1}{3}$ of the second type (B). 
Both configurations are four-coordinate, which gives the possibility to ``sneak'' into the $SiO_2$ network replacing the $4$-coordinate units $Q^4$. 

All the $Na^+$ and $Ca^{++}$ ions being now parts of $Q^3$ and $2Q^3$ configurations, we have to rescale their relative parts.

Of $100$ molecules of the window glass melt there were $75$ $SiO_2$ molecules, $15$ $Na_2O$ molecules and $10$ $CaO$ molecules.

The $30$  $Na^+ $ ions coming from the dissociation of $15$ $Na_2O$ molecules transform $30$ $Q^4$ units into $30$ $Q^3$ ones.

The $10$ $CaO$ molecules produce $10$ $2Q^3$ clusters with $20$ $Q^4$ units, leaving only $25$ $Q^4$ units unaltered. 
In a borosilicate glass with $z$ $B_2O_3$ molecules replacing the same amount of $SiO_2$, the new distribution is: 
$(0.75 - z) \; SiO_2 + 0.15 \;  Na_2O + 0.10 \;  CaO + z \; \; B_2O_3 $

\begin{figure}[hbt]
\centering 
\includegraphics[width=7.8cm, height=3.3cm]{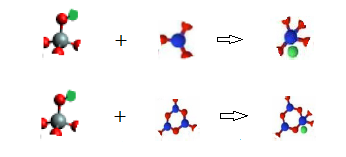}
\caption{Above: creation of $B4$ by exchange of an $Na^{+}$ ion between $Q^4$ and $B^3$; below: creation of a $Tetraborate$}
\label{fig:reactions2}
\end{figure}

Let us evaluate the impact on average rigidity $r$ per atom, coming from each structural unit. The average rigidity per atom being defined as 
$$ r = \frac{N_{\alpha} + N_{\lambda}}{N_a} - 3,$$
where $N_{\alpha}$ is the total number of angular constraints, 

$N_{\lambda}$ is the total number of linear constraints,

$N_a$ is the total number of atoms belonging to a given structural unit.  

When the average rigidity $ r < 0$, the unit is ``floppy'', i.e. underconstrained; when $ r = 0$
it is isostatic, and when $ r > 0$ it is ``rigid'', or overconstrained 

\begin{figure}[hbt]
\centering
\includegraphics[width=11cm, height=2.1cm]{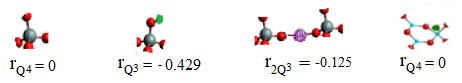}
\label{fig:Rigidity}
\caption{Local $4$-coordinate clusters in $(1-x) B_2O_3 + x Na_2O$ and their rigidity defect $r$}
\end{figure}

Let us show on a particular example how the rigidity $r$ is evaluated. Starting with the $Q^4$, we have $5$ angular constraints for the $4$-coordinate
$Si$ atom, and four linear constraints - the covalent bonds towards the Oxygens. In a random glassy $SiO_2$ network the angular constraints around oxygen bonds are broken;
therefore the total number of constraints is $N_c =9$. The number of atoms is $3$ (one $Si$ and four halves of $O$), and we get $r = N_c/N_a - 3=0$; the cluster is {\it isostatic}.
For the $Q^3$ the constraints are the same, $N_c = 9$, but the number of atoms is $3.5$ - one $Si$, one entire $O$ and three halves (the oxygens shared with other $Si$ neighbors). 
We do not count the $Na^{+}$ ion as being a part of the network. The result is $r = N_c/N_a - 3 = 9/(3.5) -3 = -0.429$. The $2Q^3$ configuration with $Ca$ bridge has $N_c = 21$,
from $11$ angular constraints, $2 \times 5$ around the $4$-fold $Si$ atoms and the stiff $Ca^{++}$ bridge in between, plus $10$ linear constraints, plus $10$ linear constraints.
The number of atoms is $8$, including the $Ca$ which here is the part of the network. As a result, we get $r = N_c/N_a -3 = 21/8 - 3 = -0.125$. The Tetraborate is isostatic, as shown 
in our previous papers on alkali-borate glasses (\cite{DMDSRKMM}, \cite{RKDSLCR}). 

In a soda-lime silicate glass, where we have $30 \% $ of $Q^3$ units, $50 \%$ of $Q^4$ and $20 \% $ of $2Q^3$ units, the overall rigidity defect (per atom) 
is $-0.253$, obtained as follows: $<r> = <N_c>/<N_a> - 3$, with $<N_c> = 0.5 \times 9 + 0.3 \times 9 + 0.2 \times 21 = 11.4$ 
and $<N_a> = 0.5 \times 3 + 0.3 \times 3.5 + 0.2 \times 8 = 4.15$. The network is underconstrained (floppy). 

During the reaction of exchange of an $Na^{+}$ ion resulting in the replacement of one $Q^3$ unit by a $B^4$ unit, the
statistical factors to be chosen are $m_A = 1, \; \; m_B = 3$. This is because the $Q^3$ presents only one dangling bond with $Na^+$ ion attached to it,
whereas a $B^3$ or a boroxol ring $B_3 O_3 {\O{}}_3$ has three places to which the $Na^+$ can be attached creating an extra oxygen bond.   
Let us apply the initial slope formula to the $T_g (z)$ curve for the soda-lime borosilicates with chemical composition given by
$(0.75-z) \; SiO_2 + 0.15 \; Na_2O + 0.10 \; CaO + z \; B_2O_3.$, displayed in Fig. $11$,
produced using the experimental data given in Table $1$, which are very close to the ``ideal'' composition

$(0.75 - z) SiO_2 + 0.15 \; Na_2O + 0.10 \; CaO + z \; B_2O_3$ 

The deviations from this formula are quite negligible, taking into account the scale and precision of the graph.
The continuous curve is the result of spline interpolation method, implemented by computer algebra system. In order to improve the visibility, it was copied with continuous
interpolation reinforced with the use of computer graphic program.

\begin{figure}[hbt]
\centering 
\includegraphics[width=13.5cm, height=6cm]{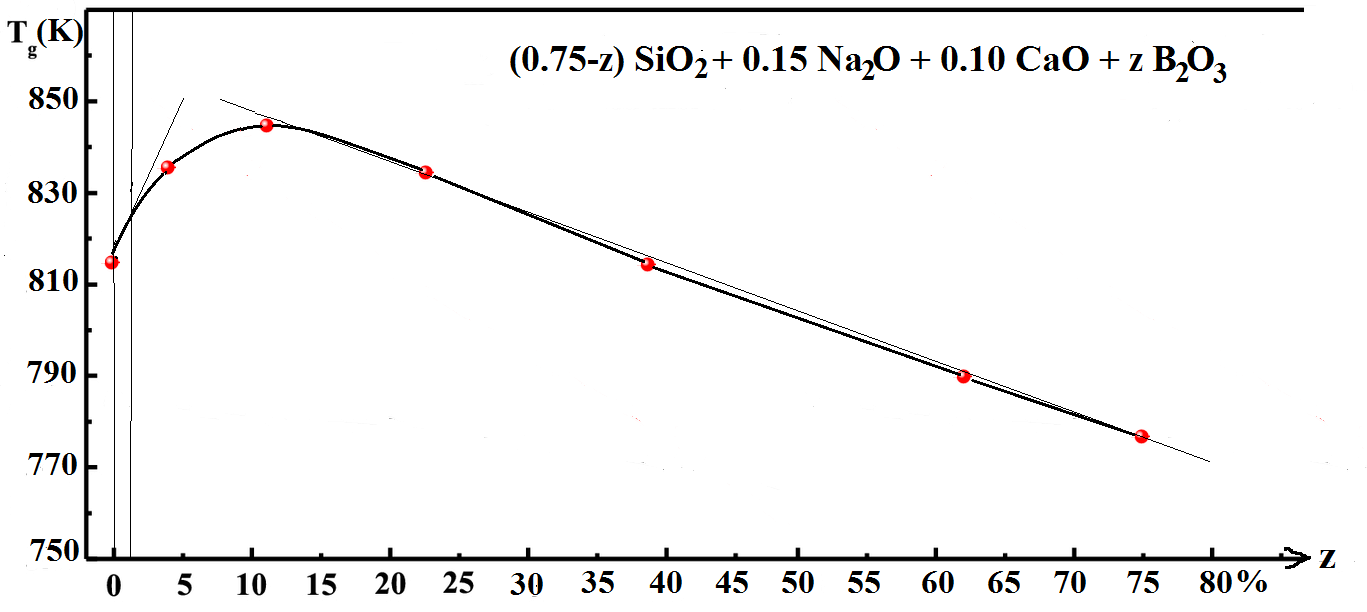}
\label{fig:BorSi2_Tg}
\caption{The curve $T_g (z)$ displays three distinct phases: firstly, from floppy to isostatic composition, $0 < z < 0.10$, 
secondly, the isostatic region $0.10 < z < 0.15$, and finally a linear behavior accompanying the steadily decreasing rigidity of the floppy network}
\end{figure}

The formula to be checked against the experimental data is 
\begin{equation}
\label{elegant2}
\left.\frac{\mathrm{d}T_{\mathrm{g}}}{\mathrm{d}z}\right|_{z=0}= p \; \frac{T_{\mathrm{g}0}}{\ln
\left(\frac{m_{\mathrm{B}}}{m_{\mathrm{A}}}\right)},
\end{equation}
where $\frac{m_B}{m_A} = 3$ and $p=2$ because ONE $B_2O_3$ molecule transforms TWO $Q^3$ items into two $B^4$ units. 

Let us check: 
$$\ln(\frac{m_B}{m_A}) = \ln 3 = 1.0986, \; \; \; T_g (z=0) = 814K,$$
therefore
$$\left.\frac{\mathrm{d}T_{\mathrm{g}}}{\mathrm{d}z}\right|_{z=0}= 2 \; \frac{T_{\mathrm{g}0}}{\ln
\left(3\right)} = 2 \; \frac{814\; {\rm K}}{1.1} = 1482 \; {\rm K} / {\rm mole}.$$
It means that at $z=0.01$, i.e. $1 \%$ of $B_2O_3$ in the melt, the glass transition temperature increases by about $15$ K.
This is in a very good agreement with the experiment, as shown on the Fig. 11.

The linear character of the curve $T_g (z)$ starts at the value of $z \simeq 0.20$, when all $Na^{+}$ and $Ca^{++}$ ions are tied with Borons,
$(50 -z)$ consisting of pure $Q^4$ structural units and $z $ boron oxide $B_2O_3$. 
The equation of the straight line coinciding with the curve is given by the following expression: 
\begin{equation}
T_g (z) = T_g (0.219) +  \frac{ T_g (0.741) - T_g (0.219) }{0.741 - 0.219}  ( z - 0.219 ).
\label{Tglinear1}
\end{equation}
or, inserting the experimental data,  
\begin{equation}
T_g(z) = 833 \; {\rm K} - \frac{833 \; {\rm K} - 775 \; {\rm K} }{0.741 - 0.219} ( z - 0.219).
\label{Tglinear2}
\end{equation}
According to formula \ref{Tlinear}, the linear character of the $T_g(z)$ curve indicates that two types of aglomerating clusters have the same
coordination number. We conclude that in this case $m_A = m_B = 4$. The first ($m_A$ )refers to the average coordination number of $Q^4$ entities dominant
in window glass former past $z \simeq 0.20$, the second ($m_B$) referring to four-coordinate alkali-borate clusters shown in Fig. 8.

\section{Conclusion}

We have shown how the stochastic agglomeration model can be successfully applied to soda-lime borosilicates. Although conceived for modelling
of simple covalent chalcogenide glasses, it conserves its predictive power also in the case of oxides, provided that one correctly identifies the
stable clusters replacing single atoms as the elementary network building blocks. 

Two most salient properties of the $T_g(z)$ curve are in agreement with the model, confirming the assumptions about the energy costs of creating
particular clusters containing a $Na^+$ ion, the potential well being deeper for a $B_4$ creation than for a $Q^3$ with one dangling bond.

The dependence of $T_g$ on the average rigidity also confirms former investigations applied to chalcogenides \cite{Naumis2006},and alkali-borates  
\cite{RKRBJPD}. However, the Gibbs-di Marzio phenomenological formula \cite{DiMarzio1958}, successfully applied to alkali-borates \cite{RKDSLCR}
did not give a good agreement with the experiment here, probably due to the more sophisticated dependence between average rigidity and distribution
of vibrational modes, more complicated than in the case of pure borate glasses.

The Volterra-type differential equation approach, successfully tested on alkali borate glasses in \cite{RKDSLCR} could be also employed here. The boroxol
rings could be given the predator role, while the $Q^3$ silicons with one dangling bond being the prey. But it would need 
much more programming skills due to the complexity of configurations. We shall adress this issue in the forthcoming publications.  

\vskip 0.4cm
\indent
\hskip 0.5cm
{\bf Acknowledgements}
\vskip 0.3cm
\indent
I gratefully acknowledge many inspiring and fruitful discussions with Natalia M. Vedishcheva and for her pertinent and helpful advices. 
Special thanks to Alex C. Hannon for numerous advices and invitation to present these results at the Cambridge 2025 SGT Conference.
I also gratefully acknowledge and thank the anonymous Referees whose critical remarks and constructive suggestions enabled the author to
substantially improve this article.

\end{document}